 \documentclass[twocolumn,times]{aastex6}
\usepackage{url,color}
\usepackage[]{natbib}

\begin{document}

\title{The Mid-Infrared Evolution of the FU Orionis Disk}

\author{
Joel D. Green\altaffilmark{1,2},
Olivia C. Jones\altaffilmark{1},
Luke D. Keller,\altaffilmark{3},
Charles A. Poteet\altaffilmark{1},
Yao-Lun Yang\altaffilmark{2},
William J. Fischer\altaffilmark{4},
Neal J. Evans, II\altaffilmark{2},
Benjamin A. Sargent\altaffilmark{1},
\& Luisa M. Rebull \altaffilmark{5}
}

\altaffiltext{1}{Space Telescope Science Institute, Baltimore, MD, USA}
\altaffiltext{2}{Dept of Astronomy, The University of Texas at Austin, Austin, TX 78712, USA}
\altaffiltext{3}{Department of Physics \& Astronomy, Ithaca College, Ithaca, NY, USA}
\altaffiltext{4}{NASA Postdoctoral Program Fellow, Goddard Space Flight Center, Greenbelt, MD, USA}
\altaffiltext{5}{IPAC, Pasadena, CA, USA}

\begin{abstract}

We present new SOFIA-FORCAST observations obtained in Feburary 2016 of the archetypal outbursting low mass young stellar object FU Orionis, and compare the continuum, solid state, and gas properties with mid-IR data obtained at the same wavelengths in 2004 with Spitzer-IRS.  In this study, we conduct the first mid-IR spectroscopic comparison of an FUor over a long time period.  Over a 12 year period, UBVR monitoring indicates that FU Orionis has continued its steady decrease in overall brightness by $\sim$ 14\%.  We find that this decrease in luminosity occurs only at wavelengths $\lesssim$ 20 $\mu$m.  In particular, the continuum short ward of the silicate emission complex at 10 $\mu$m exhibits a $\sim$ 12\% ($\sim$ 3$\sigma$) drop in flux density, but no apparent change in slope; both the Spitzer and SOFIA spectra are consistent with a 7200 K blackbody.  Additionally, the detection of water absorption is consistent with the Spitzer spectrum.  The silicate emission feature at 10 $\mu$m continues to be consistent with unprocessed grains, unchanged over 12 years.  We conclude that either the accretion rate in FU Orionis has decreased by $\sim$ 12-14\% over this time baseline, or that the inner disk has cooled, but the accretion disk remains in a superheated state outside of the innermost region.
\end{abstract}


\section{Introduction}

The low mass pre-main sequence star FU Orionis (hereafter, FU Ori) brightened by $\Delta B = 6$ mag in 1936 \citep{herbig77}, and subsequently has declined slowly ($\sim$ 0.013 mag/yr; \citealt{kenyon00}) to the present day (further examined in this work). This burst in brightness established a class of variable star (FU Orionis objects, or ``FUors''; \citealt{hartmann96a,hartmann98}) that $\sim$10 other stars have since mimicked. \citet{paczynski76} first proposed that FUors are the result of a sudden cataclysmic accretion of material from a reservoir that had built up in the circumstellar disk surrounding a young stellar object. In this model, the accretion rate rises from the typical rate for a T Tauri star ($\dot{M} \lesssim 10^{-7} M_{\odot}$ yr$^{-1}$) up to $\dot{M} \sim 10^{-4} M_{\odot}$ yr$^{-1}$, and then decays over an e-fold time of $\sim$ 10-100 yr \citep{lin85,hartmann85,bell94}.   Over the entire outburst the star could accrete $\sim$ 0.01 M$_{\odot}$ of material, roughly the entire mass of the Minimum Mass Solar Nebula or a typical T Tauri disk \citep{andrews05}.  On average, FUor outbursts should occur (or recur) $\sim$ 5 times per star formed in the local region of the Milky Way \citep{hartmann96a}; there have been $\sim$ 10 FUors observed within a distance of 1.5 kpc in the last 80 yr, compared with the star formation rate in the same volume ($\sim$ 3 per 50 yr). These bursts also modify the protoplanetary disk chemistry and require a very different model than simple magnetospheric accretion \citep[e.g.][]{green06, quanz07a, zhu07,romanova09,cieza16}, usually invoked for accreting young stars.  If FUors represent a short-duration stage that all young stars undergo, then understanding their properties is vital to models of planet formation and the evolution of protoplanetary gas-rich disks.  

The luminosity outburst is due to a sudden increase in the rate at which material accretes from the disk onto the star.  Models explain the rise in accretion rate with the development of instabilities in the disk, but they differ as to where the instabilities originate.  ``Inside-out" models are those in which the instability develops at several AU, where the lower temperatures, and therefore lower ionization fraction, render the magneto rotational instability (MRI) as an inefficient accretion mechanism \citep[e.g.][]{zhu09b,martin12}.  ``Outside-in'' models are those in which gravitationally bound fragments develop at large radii (tens of AU or more) due to infall from a protostellar envelope, then migrate inward through the disk, ultimately generating an outburst when they accrete onto the star \citep[e.g.][]{vorobyov15}.  A third option that has been posited is an external perturber -- the interaction of the disk with an exterior massive object that alters the balance of accretion and triggers the cascading flow \citep{bonnell92}, although recent studies suggest the refill rate is too slow to occur on few co-orbit timescales \citep{green16b}. The wavelength dependence of any post-outburst variability can distinguish among these classes of models.

Dust grain growth, processing, and sedimentation are tracked by mid-infrared (MIR) dust features.  Processing and grain growth in the disk can begin as early as the protostellar stage, when the disk is heavily extinguished. Most silicate features around low mass young stars are consistent with both grain growth and high crystallinity fractions \citep[e.g.][]{watson09,olofsson09}.  Because  grains  from the ISM enter the disk as amorphous, high crystalline fractions in young stars indicate that dust processing occurs early on. EX Lup, the archetype of a smaller magnitude burst from a T Tauri disk, exhibited crystalline grain production during outburst \citep{abraham09}, due to enhanced heating.  However, these crystalline grains disappeared during quiescence, likely due to X-ray amorphitization of the grains, or dust transport either radially or vertically.  In contrast, FUors show no evidence for processed grains during outburst, only grain growth \citep{quanz07a}.  No pre-outburst mid-IR spectra of FUors exist to constrain their historical grain properties, and it may be that the silicate grains are modified during the outburst.    For protoplanetary disks, the 10 and 20 $\mu$m silicate features probe optically thin dust at specific temperatures on the disk surface; as the disk heats and then cools, the radii probed by these features changes accordingly, and may evolve during the burst \citep[e.g.][]{hubbard14}.  As the disk cools, we might expect additional changes to occur, as grains are recirculated, modified, merged, or destroyed.  During an FUor, the temperature at 1 AU may approach 1000 K or even higher; this becomes comparable to the temperature to form crystalline silicates or remove volatiles, which can affect habitable zone planet formation \citep{hubbard14}.  Further out in the envelope, episodic increases in temperature can cause irreversible chemical processes, such as the observed conversion of mixed CO-CO$_2$ ice into pure CO$_2$ \citep{kimhj12b,poteet13}.

Mid-IR spectroscopic variability has been only sparsely examined, although mid-IR evolution during bursts has been modeled \citep[e.g.][]{bell94,zhu07,johnstone13}. The biggest difference between the mid-IR SED of FU Ori and that of more typical young stellar objects is contribution of the superheated inner disk.  The innermost regions of disk are expected to reach temperatures of 6000-8000 K, comparable to the stellar surface but over vastly greater emitting area (the inner radius of the disk;  \citealt{hartmann96a,hartmann98}).  Therefore the innermost disk radii are no longer coupled to any dust, which would be destroyed or ejected out to a significant fraction of an AU, and would not be an opacity source.  Thus FU Ori's SED is dominated at {\it all wavelengths} by disk emission, even in the optical/near-IR, and changes at optical/near-IR wavelengths would reflect changes in the superheated disk rather than the central protostar.

FU Ori (D $=$ 450-500 pc; \citealt{hartmann96a}) is a wide (0$\farcs$5) binary system consisting of a outbursting young stellar object modeled as a 0.3  $M_{\odot}$ with a 0.04  $M_{\odot}$ circumstellar disk \citep{zhu07} at an A$_V$ of $\sim$ 1.8 \citep{zhu10}, and a more massive young companion (M$>$ 0.5 $M_{\odot}$; \citealt{wang04,reipurth04,hales15}).  The companion (FU Ori S) is at 227 AU projected separation \citep{pueyo12}, at an A$_V$ of 10 $\pm$ 2 \citep{pueyo12}.

In this work, we present new mid-IR photometry and spectroscopy of FU Ori obtained with the Stratospheric Observatory for Infrared Astronomy (SOFIA) in 2016, and compare with previous observations with the Spitzer Space Telescope in 2004.  We find that the continuum flux density in at wavelengths shorter than 8 $\mu$m (arising from the hot inner disk) has decreased by 12-14\%, dominating the change in bolometric luminosity.  The continuum at wavelengths greater than 10 $\mu$m has decreased only 7\% within uncertainties.  Additionally, we find that the H$_2$O vapor absorption resulting from the superheated disk photosphere remains present.  Finally, we note no significant changes in the 10 $\mu$m silicate emission feature, which shows no evidence of crystallinity.  Taken together, these observations suggest that the decrease in viscous dissipation energy over this period has occurred predominantly in the innermost section of the disk, which is changing rapidly.

\section{Observations}

\setlength{\tabcolsep}{0.065in}
\renewcommand{\arraystretch}{0.55}
\begin{deluxetable}{lllrrr}
 \tablecaption{Observing Log}
 \tablehead{\\
 \colhead{Obs Mode} & \colhead{AORIDs} & \colhead{Filter} & \colhead{Int(s)} & Num & \colhead{Total(s)}} 
 \startdata
 \\
Im. & 0401461, 040146165 & F056/F253 & 60 & 1 & 184 \\
Im. & 0401462, 041046166 & F077/F315 & 60 & 1 & 184 \\
Im. & 0401463, 040146163 & F111/F348 & 147$^1$ & 1 & $\sim$ 400$^1$ \\
Grism  & 04014616 & G063 & 200 & 2 & 550$^2$ \\
Grism & 04014617 & G111 & 3600 & 5 & 9225$^3$ \\
Grism & 04014618 & G227 & 1000 & 1 & 2577 \\
IRS Stare &  3569920 & SL/SH/LH & 18 & 1 & -- \\
\\
\enddata
\tablecomments{Observations were performed over two different flights, 272 (04 Feb 2016) and 279 (18 Feb 2016).\\
$^1$ The F111 band was observed in 7 separate integrations; six were 13s on source and one was 69s on source.  We report the combined exposure time and approximate overhead. \\
$^2$ The exposure on Flight 272 exhibited $\sim$ 50\% worse S/N than the exposure on Flight 279 and was not used in the final spectrum presented in this work. \\
$^3$ The two exposures on Flight 272 exhibited 10\% higher continuum than the three exposures on Flight 279, further misaligned with photometric measurement, and were not used in the final spectrum presented in this work. \\
 }
 \label{obslog}
 \end{deluxetable}
 
 \subsection{SOFIA-FORCAST Spectroscopy}
 
We observed FU Ori with the Faint Object infraRed CAmera for the SOFIA Telescope (FORCAST; \citealt{herter12}) onboard the Stratospheric Observatory For Infrared Astronomy (SOFIA; \citealt{young12}) on two separate flights (on 2016 Feb 4 and Feb 18, flights 272 and 279 respectively); see Table \ref{obslog} for the Observing Log.  The observations were performed in ``C2N'' (2-position chop with nod) mode, using the 4$\farcs$7 slit, and the NMC (nod-match-chop) pattern in which the chops and nod are in the same direction and have the same amplitude.  In each case, an off-source calibrator was selected from an ensemble of observations from Cycles 2-4, using the observation closest in zenith angle and altitude to the science target (A. Helton, priv. comm.).  We did not use dithering.  The FORCAST grism was used in three spectral bands G063 ($\sim$ 5-8 $\mu$m) G111 ($\sim$ 9-13 $\mu$m) and G227 ($\sim$ 18-25 $\mu$m) from 5-25 $\mu$m  to achieve a spectral resolving power R $\sim$ 200.  The achieved signal-to-noise ratios for the individual grism segments were 30 (G063, measured from 5.7 to 6.3 $\mu$m), 46 (G111; measured from 11-12 $\mu$m), and 22 (G227, measured from 21-26 $\mu$m). The total observation time was 12904 seconds (3.58 hr), including imaging, which totaled approximately 0.33 hr  (see below).

The extracted SOFIA-FORCAST spectra are the best calibrated version of the Level 3 pipeline products, as identified in the SOFIA Data Cycle System.  The G227 observation appears in only one exposure/AOR, but the G063 was observed on two separate occasions (one on each of two flights), and G111 was observed in five separate exposures to obtain sufficient signal to noise ratio (three exposures on Flight 279 and two exposures on Flight 272).  Notes on each individual grism observations follow.

{\it G063.} The Flight 279 G063 spectrum shows 50\% higher noise than the Flight 272 spectrum, and we decided to exclude the low S/N data; this did not significantly affect the absolute flux density in the G063 band.  

{\it G111.} However, the two Flight 272 G111 spectra are both 10\% higher in flux density than the three Flight 279 spectra, and we excluded the {\it higher flux density} G111 spectra from the analysis.  Additionally, the G111 spectrum includes several bands dominated by atmospheric ozone.  Although the SOFIA pipeline attempts to correct for this using off-positions, the telluric correction cannot reliably remove the ozone feature because the ozone layer is inhomogeneous and the SOFIA beam tracks in and out of higher density regions on timescales shorter than the background removal process.  Thus we masked this section of spectrum ($\sim$ 9.3-10.05 $\mu$m) and did not consider it in this work.

{\it G227.} The G227 spectrum showed the highest overall uncertainty.  The spectrum was re-reduced using a custom pipeline (W. Vacca, priv. comm.).  A preliminary version of a new asteroid response curve along with a different background region than that used for the original pipeline extraction was applied.  The result shifted the G227 spectrum down $\sim$ 7\% from the default, in between the three longest wavelength SOFIA photometric bands.

\subsection{Spitzer-IRS Spectroscopy}

FU Ori was observed by the InfraRed Spectrograph \citep{houck04} onboard the Spitzer Space Telescope \citep{werner04} as part of the IRS\_Disks guaranteed time observing program \citep{houck04} on 4 Mar 2004 (Table \ref{obslog}).  The observations (AORID 3569920) were  performed in Staring Mode, with two nods spaced by one third of the slit width, and off-positions acquired only for the Short-Low (SL; 5.3-14 $\mu$m, R $\sim$ 60-120) modules.  The Short-High (SH; 10-20 $\mu$m, R $\sim$ 600) and Long-High (LH; 20-38 $\mu$m; R $\sim$ 600) modules did not include off-positions, and were extracted as full aperture products \citep{green06}.  To confirm that the IRS spectral reduction was consistent with the most recent calibration, we retrieved the data from the Cornell Atlas of Spitzer IRS Sources (CASSIS LRv7 and HRv1; \citealt{lebout11,lebout15}).  The spectrum was then scaled to SL, adjusting SH fluxes upward by 2\% and LH fluxes downward by 3\%.  Details on the original data reduction are published in \citet{green06} using the SMART package \citep{higdon04}.

The absolute calibration uncertainty of Spitzer-IRS is typically as large as 10\%.  However, the point-to-point (relative) uncertainty is $\sim$ 3\%.  After cross-calibrating SL with SH and LH as noted above, the absolute flux uncertainty is only the uncertainty in SL, which is 10\%; however, the IRAC photometry, which covers the same bands, has an uncertainty floor of only 3\% \citep{carey12}.  Thus we take the final uncertainty in absolute flux calibration for Spitzer to be 3\%.

\subsection{Photometry}

\begin{figure*}
\centering
\includegraphics[scale=0.3,trim={0 0 0 0},clip]{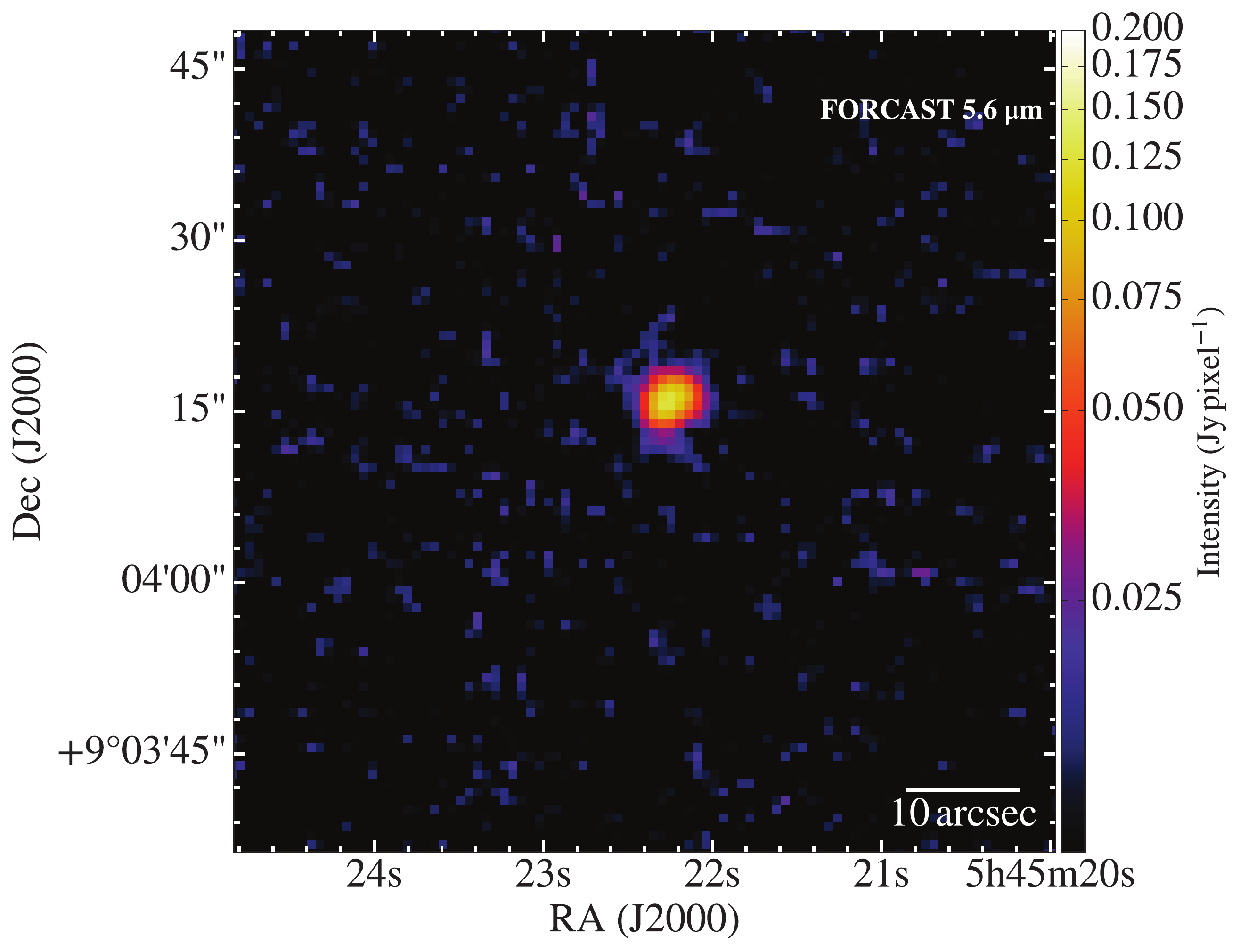}
\includegraphics[scale=0.3,trim={0 0 0 0},clip]{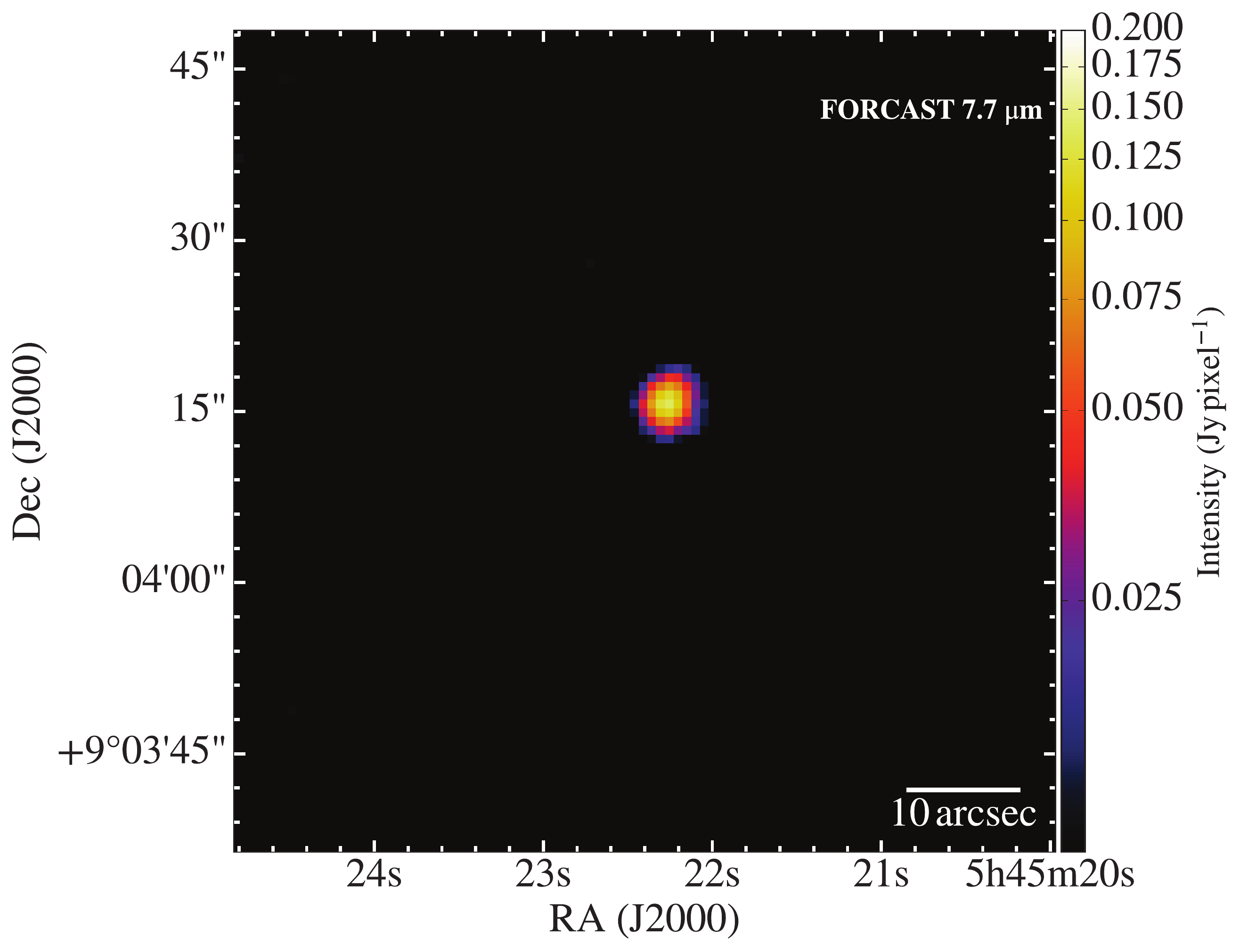} \\
\includegraphics[scale=0.3,trim={0 0 0 0},clip]{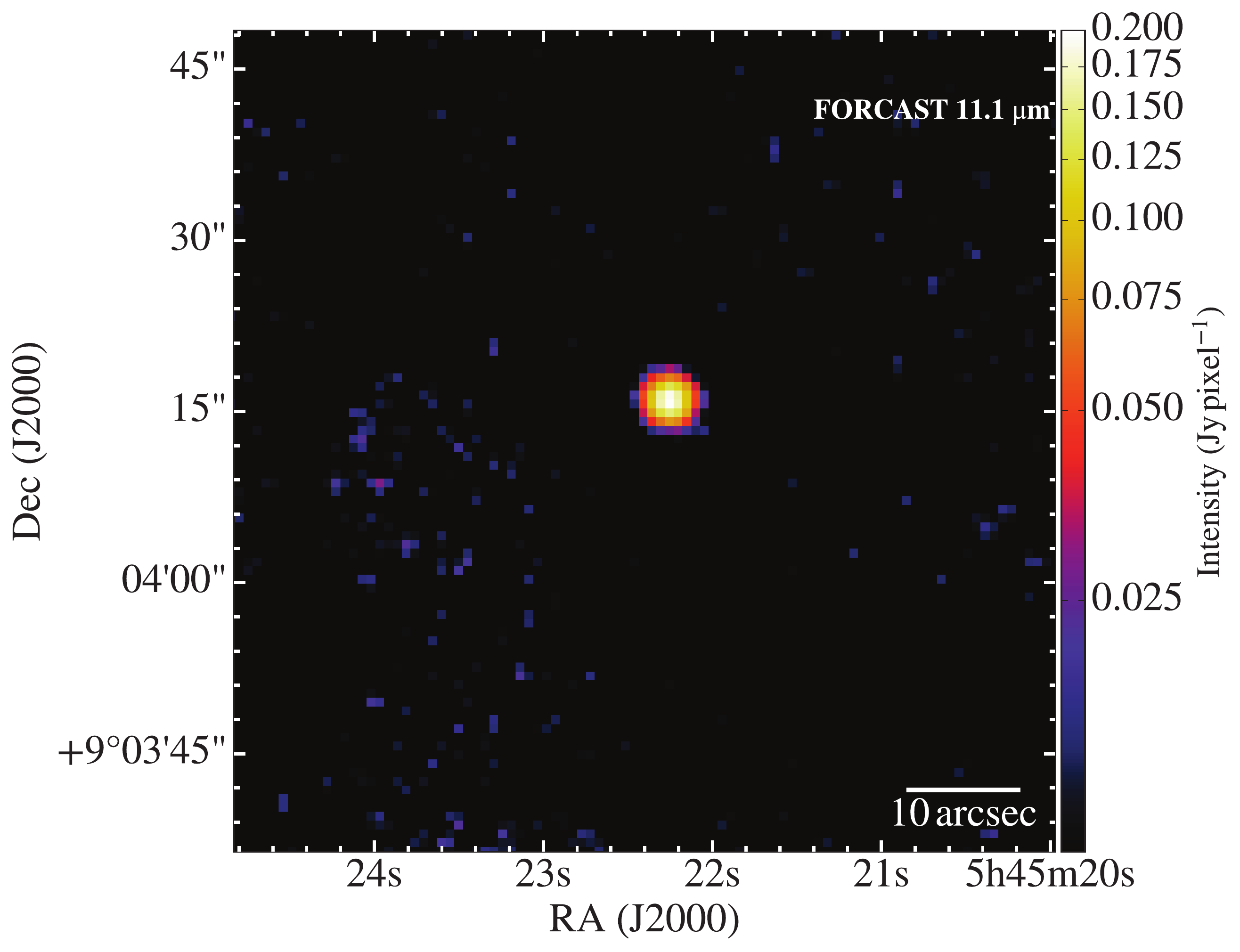}
\includegraphics[scale=0.3,trim={0 0 0 0},clip]{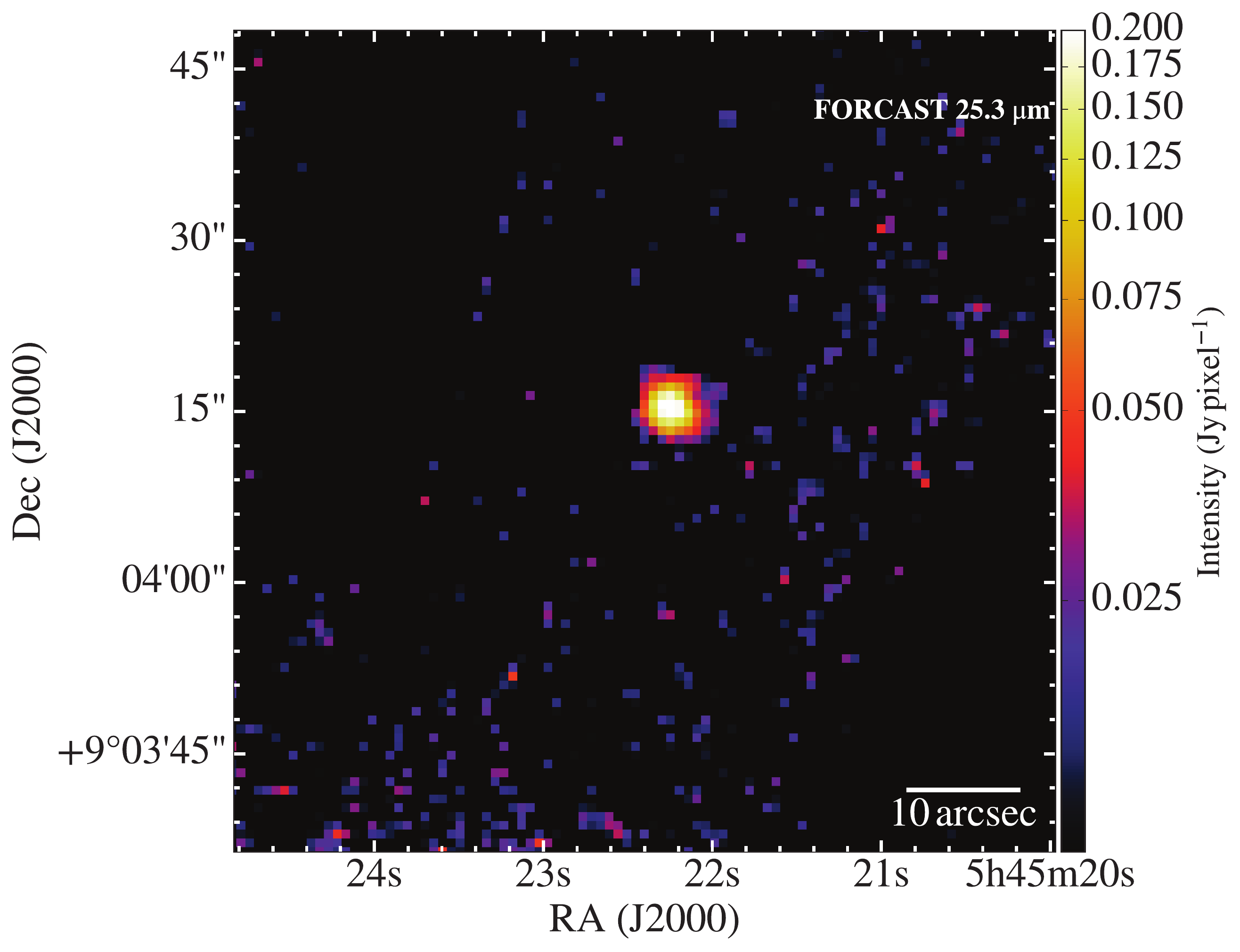} \\
\includegraphics[scale=0.3,trim={0 0 0 0},clip]{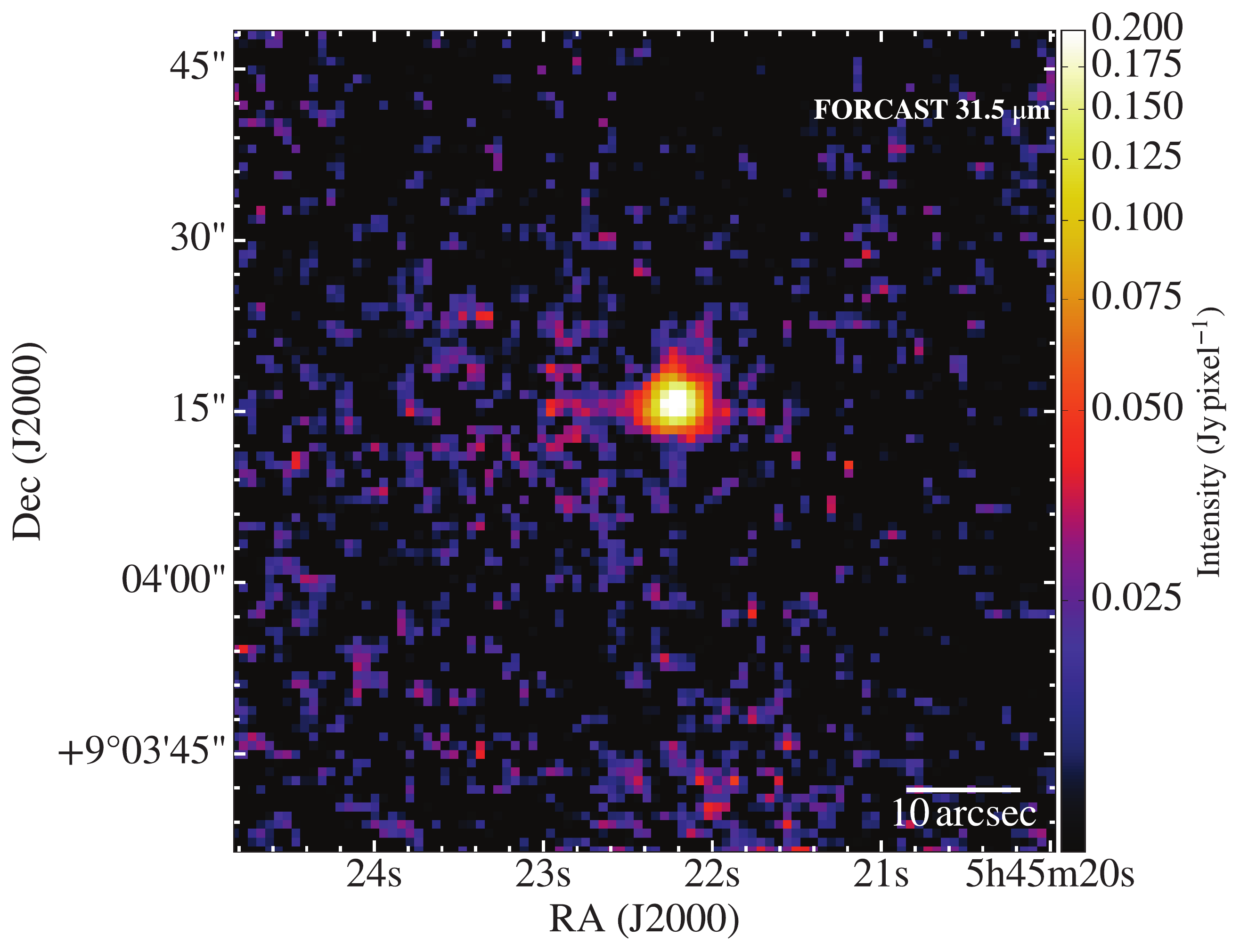}
\includegraphics[scale=0.3,trim={0 0 0 0},clip]{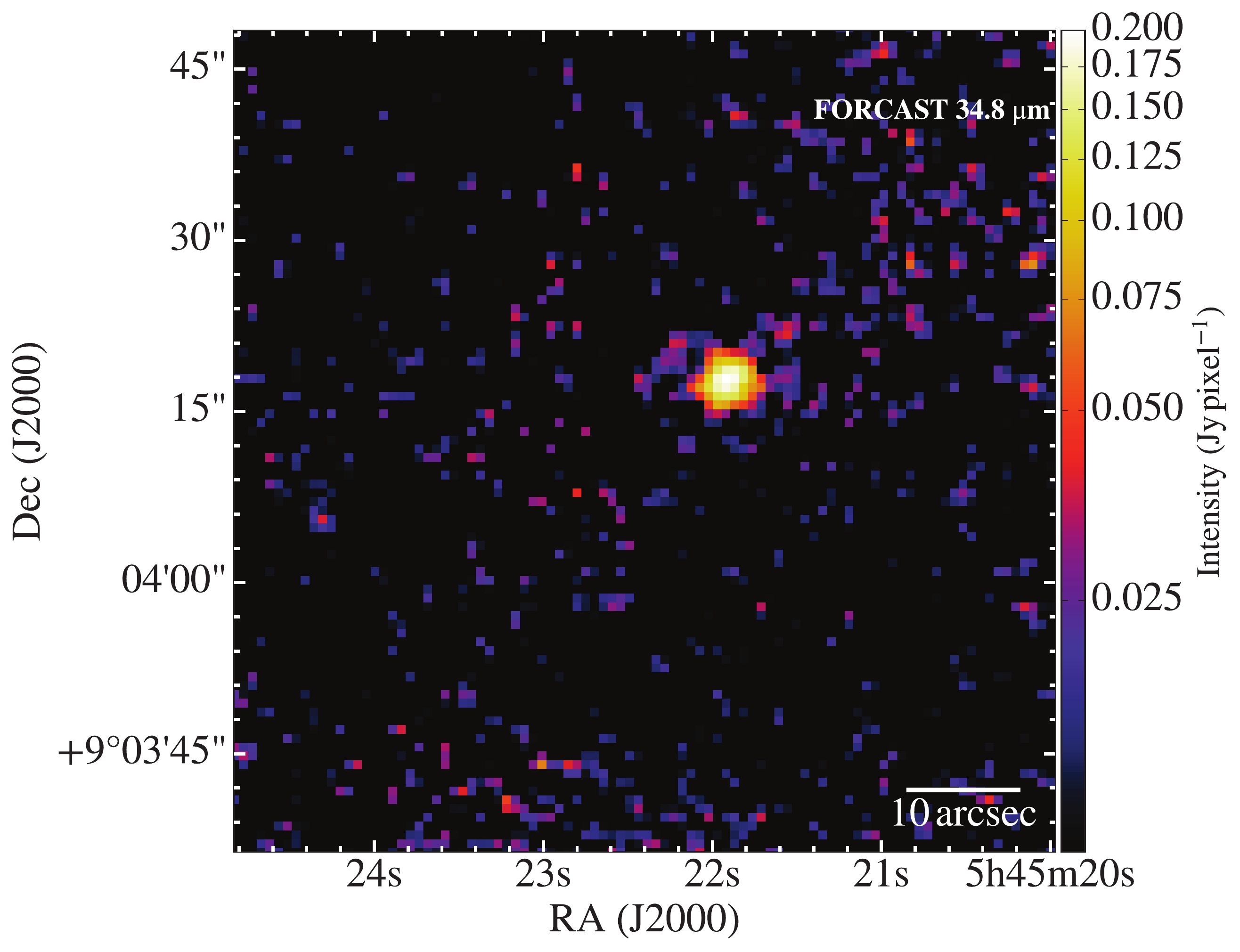} \\
\figcaption{\label{fig:image}  5.6, 7.7, 11.1, 25.3, 31.5 and 34.8 $\mu$m broadband images of FU Ori from SOFIA-FORCAST, at arcmin scales to show the surrounding background, at the same absolute intensity scale.  The apparent difference in centroid position is due to differences in the WCS pipeline processing.}
\end{figure*}

We acquired contemporaneous FORCAST imaging in F056/F253, F077/F315, and F111/F348 in order to calibrate the spectral flux densities, characterize extended emission, and provide 2016 continuum levels from 25-35 $\mu$m. All imaging observations were acquired during Flight 272.  A signal-to-noise ratio of $>$ 10 on-source was achieved in exposures of 60 seconds or less in all six bands on each source. The SOFIA-FORCAST photometry was obtained from the Level 3 pipeline data products.  The measured image quality of the SOFIA imaging ranges between 2$\farcs$7 and 3$\arcsec$ for F056, F077, and F111, and ranges between 3$\farcs$4 and 3$\farcs$6 for F253, F315, and F348; FORCAST only achieves the diffraction limit $<$ 34 $\mu$m due to telescope jitter (SOFIA Observer's Handbook, Figure 7-2). For consistency, the photometry were extracted uniformly using aperture photometry, in a 10$\arcsec$ source annulus with a sky annulus 20-40$\arcsec$ away from the source.  The F111 photometry is a weighted average of seven observations (one longer exposure was 69s on-source, and the other six short exposures were 13s on-source).  The average flux density of the six short exposures is within 1\% of the flux density measured from the long exposure.  Images for each band are shown in Figure \ref{fig:image}.  

There also appears to be some structure in the background emission, making the extraction less certain.   However, the source appears pointlike in all bands.  The absolute flux uncertainty of the level 3 G063 and G227 grism products is 5\%, and the G111 product is 10\% \citep{herter13}.  However, all of the photometric flux densities  are within 3\% of the level 3 spectroscopy.  Thus we take the combined uncertainty of the absolute flux calibration between Spitzer and SOFIA (in quadrature) to be $\sim$ 4.2\%. The background ``structure'' in some of the imaging appears significant at first glance; however, the WISE band 4 image (next section) shows no sign of structure.   A more detailed extraction of F253 using a flatfield model (J. DeBuizer, priv. comm.) indicated that the background ``structure'' was due to flatfielding.

\subsection{Photometry from Earlier Epochs}

In this work we compare photometry close in time to the 2004 IRS observations with those close to the 2016 FORCAST observations (American Association of Variable Star Observers/AAVSO BVR from 2016, and FORCAST imaging from this work).   Additional photometry is presented, including UBVR photometry previously presented in \citet{green06} (ca. 2004); 2MASS JHK (ca. 1998; \citealt{skrutskie06});  WISE (all four bands; ca. 2010; \citealt{cutri13});  and Spitzer-IRAC bands 2 and 3 (ca. 2006). The WISE W2 band appears anomalously high (perhaps  due to uncertainties related to emission from H$_2$ S(9) at 4.6 $\mu$m).  Due to these uncertainties, we find no convincing detection of background dust emission. 

The Spitzer-IRAC photometry was observed but saturated beyond recovery in Band 1 and 4.  Band 2 (4.5 $\mu$m) and 3 (5.8 $\mu$m) were extracted as Spitzer Enhanced Imaging Products (SEIP\footnote{http://irsa.ipac.caltech.edu/data/SPITZER/Enhanced/SEIP/overview.html}).  We note that the FWHM is 1$\farcs$72 (4.5 $\mu$m) and 1$\farcs$88 (5.8 $\mu$m), as indicated by the IRAC Instrument Handbook (Table 2.1). The SEIP provides two extracted aperture sizes, of diameters 3$\farcs$8 and 5$\farcs$8.    The 5$\farcs$8 diameter extraction window is a much better match with the IRS data, although in contrast to the similar wavelength WISE Band 2, the IRAC Band 2 point is {\it low} by 20\% compared to models (possibly due to partial saturation).  We note that if this difference between IRAC (2006) and WISE (2010) were due to intrinsic luminosity changes in the FU Ori disk, we would expect the WISE band to be the lower of the two, but we find the reverse.  For these reasons, we exclude both points from the analysis.

\setlength{\tabcolsep}{0.06 in}
\begin{deluxetable}{cccccc}
\centering
\tablecolumns{6}
 \tablecaption{UBVR 2004-2016 Photometry} 
 \tablehead{ \\
 \colhead{JD} & \colhead{Calendar Date} & \colhead{Mag} & \colhead{$\delta$Mag} & \colhead{Filter} & \colhead{Observer}}
 \startdata
 \\
2453278.46 & 2004 Sep. 29 & 11.697 & 0.04 & U & Ibrahimov \\
2453280.45 & 2004 Oct. 1 & 11.706 & 0.04 & U & Ibrahimov \\
2453281.472 & 2004 Oct. 2 & 11.787 & 0.04 & U & Ibrahimov \\
2453283.481 & 2004 Oct. 4 & 11.788 & 0.04 & U & Ibrahimov \\
2453284.483 & 2004 Oct. 5 & 11.772 & 0.04 & U & Ibrahimov \\
2453299.491 & 2004 Oct. 20 & 11.809 & 0.04 & U & Ibrahimov \\
2453278.46 & 2004 Sep. 29 & 11.015 & 0.02 & B & Ibrahimov \\
2453280.45 & 2004 Oct. 1 & 10.971 & 0.02 & B & Ibrahimov \\
2453281.472 & 2004 Oct. 2 & 10.988 & 0.02 & B & Ibrahimov \\
2453283.481 & 2004 Oct. 4 & 11.031 & 0.02 & B & Ibrahimov \\
2453284.483 & 2004 Oct. 5 & 11.024 & 0.02 & B & Ibrahimov \\
2453299.491 & 2004 Oct. 20 & 11.022 & 0.02 & B & Ibrahimov \\
2453309.476 & 2004 Oct. 30 & 11.009 & 0.02 & B & Ibrahimov \\
2453310.46 & 2004 Oct. 31 & 10.999 & 0.02 & B & Ibrahimov \\
2453311.499 & 2004 Nov. 1 & 11.025 & 0.02 & B & Ibrahimov \\
2457447.283 & 2016 Feb. 28 & 11.122 & 0.087 & B  & DUBF \\
2457447.283 & 2016 Feb. 28  & 11.2 & 0.08 & B  & DUBF \\
2457447.283 & 2016 Feb. 28  & 11.095 & 0.086 & B  & DUBF \\
2457447.284 & 2016 Feb. 28 & 11.134 & 0.081 & B & DUBF \\
2457447.284 & 2016 Feb. 28  & 11.078 & 0.094 & B & DUBF \\
2457458.345 & 2016 Mar. 10  & 11.191 & 0.067 & B & DUBF \\
2457458.345 & 2016 Mar. 10  & 11.209 & 0.066 & B & DUBF \\
2457458.346 & 2016 Mar. 10  & 11.143 & 0.069 & B & DUBF \\
2457458.346 & 2016 Mar. 10  & 11.148 & 0.067 & B & DUBF \\
2457461.302 & 2016 Mar. 13  & 11.187 & 0.05 & B & DUBF \\
2457461.302 & 2016 Mar. 13 & 11.272 & 0.045 & B & DUBF \\
2457461.302 & 2016 Mar. 13  & 11.186 & 0.051 & B & DUBF \\
2457461.303 & 2016 Mar. 13  & 11.162 & 0.053 & B & DUBF \\
2457461.303 & 2016 Mar. 13  & 11.179 & 0.048 & B & DUBF \\
2457465.292 & 2016 Mar. 17  & 11.249 & 0.114 & B & DUBF \\
2457465.293 & 2016 Mar. 17  & 11.188 & 0.119 & B & DUBF \\
2457465.293 & 2016 Mar. 17  & 11.173 & 0.119 & B & DUBF \\
2457465.293 & 2016 Mar. 17  & 11.32 & 0.101 & B & DUBF \\
2457470.349 & 2016 Mar. 22  & 11.092 & 0.069 & B & DUBF \\
2457470.349 & 2016 Mar. 22  & 11.28 & 0.056 & B & DUBF \\
2457470.35 & 2016 Mar. 22  & 11.148 & 0.063 & B & DUBF \\
2457470.35 & 2016 Mar. 22  & 11.169 & 0.063 & B & DUBF \\
2457470.35 & 2016 Mar. 22  & 11.35 & 0.054 & B & DUBF \\
2457479.36 & 2016 Mar. 31  & 11.177 & 0.072 & B & DUBF \\
2457479.36 & 2016 Mar. 31  & 11.308 & 0.074 & B & DUBF \\
2457479.36 & 2016 Mar. 31  & 11.24 & 0.069 & B & DUBF \\
2457479.36 & 2016 Mar. 31  & 11.197 & 0.07 & B & DUBF \\
2457479.361 & 2016 Mar. 31  & 11.026 & 0.085 & B & DUBF \\
2457480.321 & 2016 Apr. 01  & 11.171 & 0.078 & B & DUBF \\
2457480.321 & 2016 Apr. 01  & 11.005 & 0.096 & B & DUBF \\
2457480.321 & 2016 Apr. 01  & 11.206 & 0.077 & B & DUBF \\
2457480.321 & 2016 Apr. 01  & 11.092 & 0.087 & B & DUBF \\
2457480.322 & 2016 Apr. 01  & 11.191 & 0.074 & B & DUBF \\
2453278.46 & 2004 Sep. 29	 &8.485 & 0.015 & R & Ibrahimov \\
2453280.45 & 2004 Oct. 1 & 8.47 & 0.015 & R & Ibrahimov \\
2453281.472 & 2004 Oct. 2 & 8.466 & 0.015 & R & Ibrahimov \\
2453283.481 & 2004 Oct. 4 & 8.488 & 0.015 & R & Ibrahimov \\
2453284.483 & 2004 Oct. 5   & 8.504 & 0.015 & R & Ibrahimov \\
2453299.491 & 2004 Oct. 20 & 8.493 & 0.015 & R & Ibrahimov \\
2453309.476 & 2004 Oct. 30 & 8.483 & 0.015 & R & Ibrahimov \\
2453310.46 & 2004 Oct. 31 & 8.486 & 0.015 & R & Ibrahimov \\
2453311.499 & 2004 Nov. 1 & 8.477 & 0.015 & R & Ibrahimov \\
2457448.274 & 2016 Feb. 29  & 8.447 & 0.012 & R  & DUBF \\
2457448.274 & 2016 Feb. 29  & 8.434 & 0.012 & R  & DUBF \\
2457448.274 & 2016 Feb. 29  & 8.429 & 0.012 & R  & DUBF \\
2457448.274 & 2016 Feb. 29  & 8.451 & 0.012 & R  & DUBF \\
2457448.274 & 2016 Feb. 29  & 8.428 & 0.012 & R  & DUBF \\
2457451.324 & 2016 Mar. 03  & 8.445 & 0.038 & R  & DUBF \\
2457451.324 & 2016 Mar. 03  & 8.413 & 0.038 & R  & DUBF \\
2457451.325 & 2016 Mar. 03  & 8.469 & 0.029 & R  & DUBF \\
2457451.325 & 2016 Mar. 03  & 8.498 & 0.023 & R  & DUBF \\
2457451.325 & 2016 Mar. 03  & 8.473 & 0.024 & R  & DUBF \\
2457458.346 & 2016 Mar. 10  & 8.542 & 0.012 & R  & DUBF \\
2457458.346 & 2016 Mar. 10  & 8.546 & 0.012 & R  & DUBF \\
2457458.347 & 2016 Mar. 10  & 8.558 & 0.012 & R  & DUBF \\
2457458.347 & 2016 Mar. 10  & 8.548 & 0.012 & R  & DUBF \\
2457458.347 & 2016 Mar. 10  & 8.557 & 0.012 & R  & DUBF \\
2457461.303 & 2016 Mar. 13  & 8.525 & 0.01 & R  & DUBF \\
2457461.303 & 2016 Mar. 13  & 8.54 & 0.01 & R  & DUBF \\
2457461.304 & 2016 Mar. 13  & 8.516 & 0.011 & R  & DUBF \\
2457461.304 & 2016 Mar. 13  & 8.543 & 0.01 & R  & DUBF \\
2457461.304 & 2016 Mar. 13  & 8.53 & 0.01 & R  & DUBF \\
2457464.315 & 2016 Mar. 16  & 8.588 & 0.017 & R  & DUBF \\
2457464.315 & 2016 Mar. 16  & 8.601 & 0.018 & R  & DUBF \\
2457464.316 & 2016 Mar. 16  & 8.557 & 0.018 & R  & DUBF \\
2457464.316 & 2016 Mar. 16  & 8.587 & 0.017 & R  & DUBF \\
2457464.316 & 2016 Mar. 16  & 8.593 & 0.017 & R  & DUBF \\
2457465.294 & 2016 Mar. 17  & 8.528 & 0.015 & R  & DUBF \\
2457465.294 & 2016 Mar. 17  & 8.579 & 0.016 & R  & DUBF \\
2457465.294 & 2016 Mar. 17  & 8.599 & 0.016 & R  & DUBF \\
2457465.294 & 2016 Mar. 17  & 8.578 & 0.015 & R  & DUBF \\
2457465.294 & 2016 Mar. 17  & 8.563 & 0.015 & R  & DUBF \\
2457470.35 & 2016 Mar. 22  & 8.554 & 0.012 & R  & DUBF \\
2457470.351 & 2016 Mar. 22  & 8.534 & 0.013 & R  & DUBF \\
2457470.351 & 2016 Mar. 22  & 8.543 & 0.012 & R  & DUBF \\
2457470.351 & 2016 Mar. 22  & 8.556 & 0.013 & R  & DUBF \\
2457470.351 & 2016 Mar. 22  & 8.543 & 0.012 & R  & DUBF \\
2457479.361 & 2016 Mar. 31  & 8.514 & 0.011 & R  & DUBF \\
2457479.361 & 2016 Mar. 31  & 8.501 & 0.01 & R  & DUBF \\
2457479.361 & 2016 Mar. 31  & 8.512 & 0.011 & R  & DUBF \\
2457479.362 & 2016 Mar. 31  & 8.477 & 0.011 & R  & DUBF \\
2457479.362 & 2016 Mar. 31  & 8.46 & 0.011 & R  & DUBF \\
2457480.322 & 2016 Apr. 01  & 8.487 & 0.01 & R  & DUBF \\
2457480.322 & 2016 Apr. 01  & 8.502 & 0.009 & R  & DUBF \\
2457480.322 & 2016 Apr. 01  & 8.493 & 0.009 & R  & DUBF \\
2457480.322 & 2016 Apr. 01  & 8.505 & 0.01 & R  & DUBF \\
2457480.322 & 2016 Apr. 01  & 8.483 & 0.01 & R  & DUBF \\
2457484.309 & 2016 Apr. 05  & 8.448 & 0.012 & R  & DUBF \\
2457484.309 & 2016 Apr. 05  & 8.456 & 0.011 & R  & DUBF \\
2457484.309 & 2016 Apr. 05  & 8.43 & 0.012 & R  & DUBF \\
2457484.309 & 2016 Apr. 05  & 8.482 & 0.012 & R  & DUBF \\
2457484.309 & 2016 Apr. 05  & 8.466 & 0.011 & R  & DUBF \\
2453278.46 & 2004 Sep. 29  & 9.654 & 0.02 & V & Ibrahimov \\
2453280.45 & 2004 Oct. 1 & 9.629 & 0.02 & V & Ibrahimov \\
2453281.472 & 2004 Oct. 2 & 9.639 & 0.02 & V & Ibrahimov \\
2453283.481 & 2004 Oct. 4 & 9.682 & 0.02 & V & Ibrahimov \\
2453284.483 & 2004 Oct. 5 & 9.677 & 0.02 & V & Ibrahimov \\
2453299.491 & 2004 Oct. 20 & 9.676 & 0.02 & V & Ibrahimov \\
2453309.476 & 2004 Oct. 30 & 9.659 & 0.02 & V & Ibrahimov \\
2453310.46 & 2004 Oct. 31 & 9.661 & 0.02 & V & Ibrahimov \\
2453311.499 & 2004 Nov. 1 & 9.652 & 0.02 & V & Ibrahimov \\
2457447.282 & 2016 Feb. 28  & 9.746 & 0.016 & V  & DUBF \\
2457447.282 & 2016 Feb. 28  & 9.79 & 0.015 & V  & DUBF \\
2457447.283 & 2016 Feb. 28  & 9.781 & 0.015 & V  & DUBF \\
2457447.283 & 2016 Feb. 28  & 9.771 & 0.015 & V  & DUBF \\
2457447.283 & 2016 Feb. 28  & 9.775 & 0.016 & V  & DUBF \\
2457448.272 & 2016 Feb. 29  & 9.796 & 0.018 & V  & DUBF \\
2457448.273 & 2016 Feb. 29  & 9.819 & 0.017 & V  & DUBF \\
2457448.273 & 2016 Feb. 29  & 9.783 & 0.018 & V  & DUBF \\
2457448.273 & 2016 Feb. 29  & 9.811 & 0.018 & V  & DUBF \\
2457448.273 & 2016 Feb. 29  & 9.738 & 0.018 & V  & DUBF \\
2457451.322 & 2016 Mar. 03  & 9.822 & 0.046 & V  & DUBF \\
2457451.322 & 2016 Mar. 03  & 9.788 & 0.04 & V  & DUBF \\
2457451.323 & 2016 Mar. 03  & 9.803 & 0.035 & V  & DUBF \\
2457451.323 & 2016 Mar. 03  & 9.855 & 0.031 & V  & DUBF \\
2457451.323 & 2016 Mar. 03  & 9.838 & 0.03 & V  & DUBF \\
2457458.344 & 2016 Mar. 10  & 9.856 & 0.019 & V  & DUBF \\
2457458.344 & 2016 Mar. 10  & 9.838 & 0.02 & V  & DUBF \\
2457458.344 & 2016 Mar. 10  & 9.867 & 0.018 & V  & DUBF \\
2457458.345 & 2016 Mar. 10  & 9.889 & 0.018 & V  & DUBF \\
2457458.345 & 2016 Mar. 10  & 9.846 & 0.019 & V  & DUBF \\
2457461.301 & 2016 Mar. 13  & 9.88 & 0.014 & V  & DUBF \\
2457461.301 & 2016 Mar. 13  & 9.887 & 0.014 & V  & DUBF \\
2457461.301 & 2016 Mar. 13.  & 9.89 & 0.013 & V  & DUBF \\
2457461.302 & 2016 Mar. 13  & 9.888 & 0.014 & V  & DUBF \\
2457461.302 & 2016 Mar. 13  & 9.903 & 0.014 & V  & DUBF \\
2457464.313 & 2016 Mar. 16  & 9.899 & 0.032 & V  & DUBF \\
2457464.313 & 2016 Mar. 16  & 9.912 & 0.031 & V  & DUBF \\
2457464.314 & 2016 Mar. 16  & 9.875 & 0.031 & V  & DUBF \\
2457464.314 & 2016 Mar. 16  & 9.913 & 0.031 & V  & DUBF \\
2457464.314 & 2016 Mar. 16  & 9.944 & 0.03 & V  & DUBF \\
2457465.291 & 2016 Mar. 17  & 9.902 & 0.03 & V  & DUBF \\
2457465.291 & 2016 Mar. 17  & 9.926 & 0.027 & V  & DUBF \\
2457465.291 & 2016 Mar. 17  & 9.926 & 0.027 & V  & DUBF \\
2457465.292 & 2016 Mar. 17  & 9.976 & 0.026 & V  & DUBF \\
2457465.292 & 2016 Mar. 17  & 9.917 & 0.028 & V  & DUBF \\
2457470.348 & 2016 Mar. 22  & 9.912 & 0.018 & V  & DUBF \\
2457470.348 & 2016 Mar. 22  & 9.889 & 0.018 & V  & DUBF \\
2457470.348 & 2016 Mar. 22  & 9.861 & 0.018 & V  & DUBF \\
2457470.348 & 2016 Mar. 22  & 9.878 & 0.018 & V  & DUBF \\
2457470.349 & 2016 Mar. 22  & 9.905 & 0.017 & V  & DUBF \\
2457479.358 & 2016 Mar. 31  & 9.806 & 0.028 & V  & DUBF \\
2457479.358 & 2016 Mar. 31  & 9.858 & 0.026 & V  & DUBF \\
2457479.359 & 2016 Mar. 31  & 9.778 & 0.029 & V  & DUBF \\
2457479.359 & 2016 Mar. 31  & 9.802 & 0.03  & V  & DUBF \\
2457479.359 & 2016 Mar. 31  & 9.825 & 0.027 & V  & DUBF \\
2457480.32 & 2016 Apr. 01  & 9.796 & 0.024 & V  & DUBF \\
2457480.32 & 2016 Apr. 01  & 9.832 & 0.022 & V  & DUBF \\
2457480.32 & 2016 Apr. 01  & 9.794 & 0.024 & V  & DUBF \\
2457480.32 & 2016 Apr. 01  & 9.822 & 0.024 & V  & DUBF \\
2457480.32 & 2016 Apr. 01  & 9.795 & 0.025 & V  & DUBF \\
2457484.307 & 2016 Apr. 05  & 9.855 & 0.038 & V  & DUBF \\
2457484.307 & 2016 Apr. 05  & 9.74 & 0.042 & V  & DUBF \\
2457484.307 & 2016 Apr. 05  & 9.843 & 0.036 & V  & DUBF \\
2457484.307 & 2016 Apr. 05  & 9.761 & 0.037 & V  & DUBF \\
2457484.307 & 2016 Apr. 05  & 9.807 & 0.036 & V  & DUBF \\
\enddata
\tablecomments{UBVR Photometry from 2004-2016 used in this work. }
 \label{photlog}
 \end{deluxetable}

\begin{figure*}
\centering
\includegraphics[scale=0.95,trim={0 0 0 0},clip]{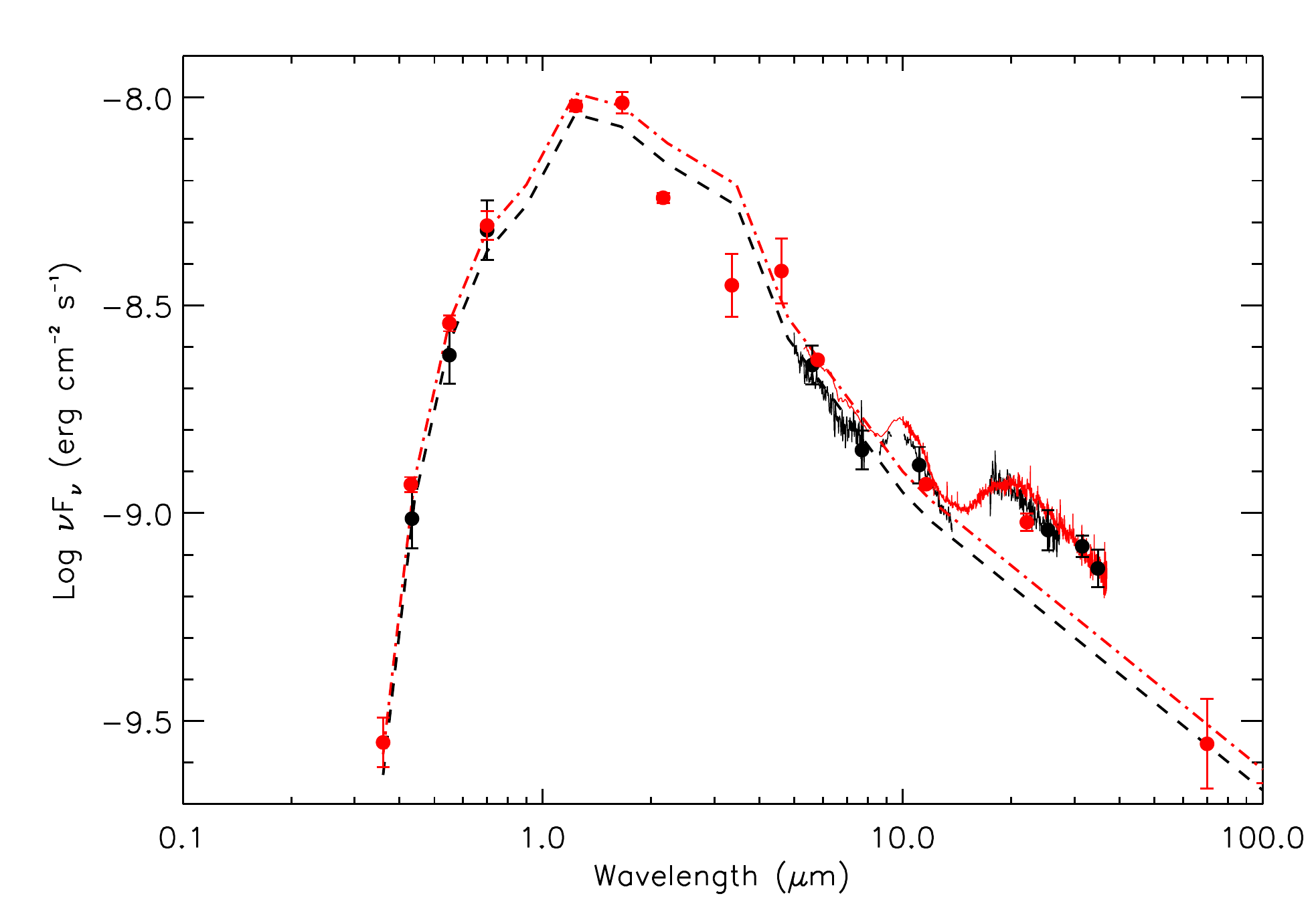}
\figcaption{\label{fig:fit} 0.3-38 $\mu$m SED of FU Ori, highlighting the decrease in luminosity between 2004 and 2016.  The (2004) Spitzer-IRS spectroscopy (red line) and pre-2011 photometry (red squares; see text for details) shortward of 8 $\mu$m are fit by a dereddened (E(B-V) $=$ 0.7; \citealt{mathis90}) 7200 K modified blackbody (\citealt{kenyon88,kenyon95}; red dot-dash line); the model is scaled assuming a distance of 450 pc to FU Ori.  The (2016) SOFIA-FORCAST spectroscopy (black line) and photometry (black circles), as well as 2016 BVR AAVSO photometry, are fit by the same 7200 K modified blackbody curve but reduced in intensity by 12\% (black dashed line). }
\end{figure*}

\begin{figure*}
\centering
\includegraphics[scale=0.75,trim={0 0 0 0},clip]{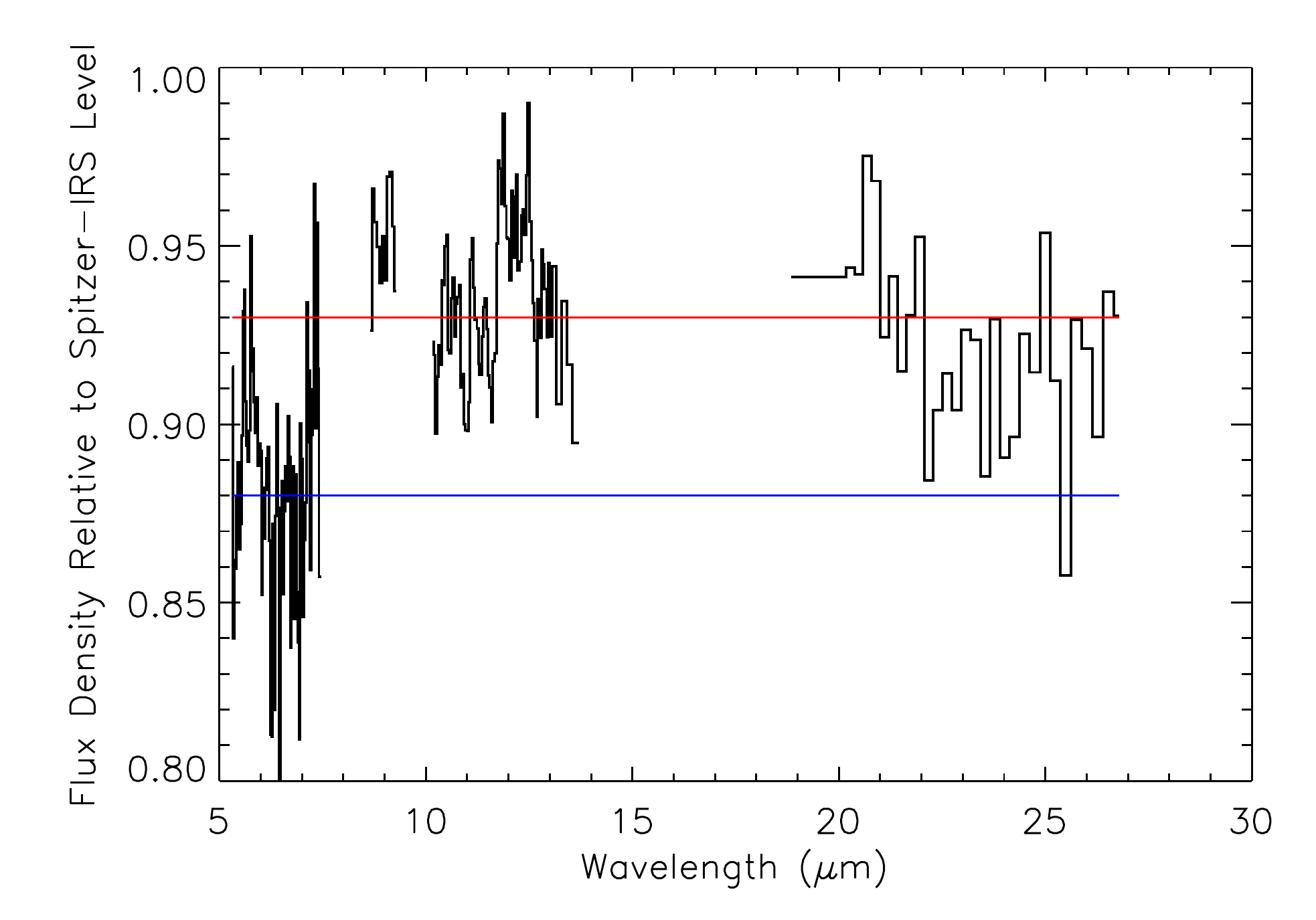}
\figcaption{\label{ratio} Ratio of the flux density of the SOFIA-FORCAST spectrum of FU Ori (2016) divided by the Spitzer-IRS spectrum (2004). For comparison, the blue line indicates a decrease of 12\%, and the red line indicates a decrease of 7\%.}
\end{figure*}

\begin{figure*}
\centering
\includegraphics[scale=0.75,trim={0 0 0 0},clip]{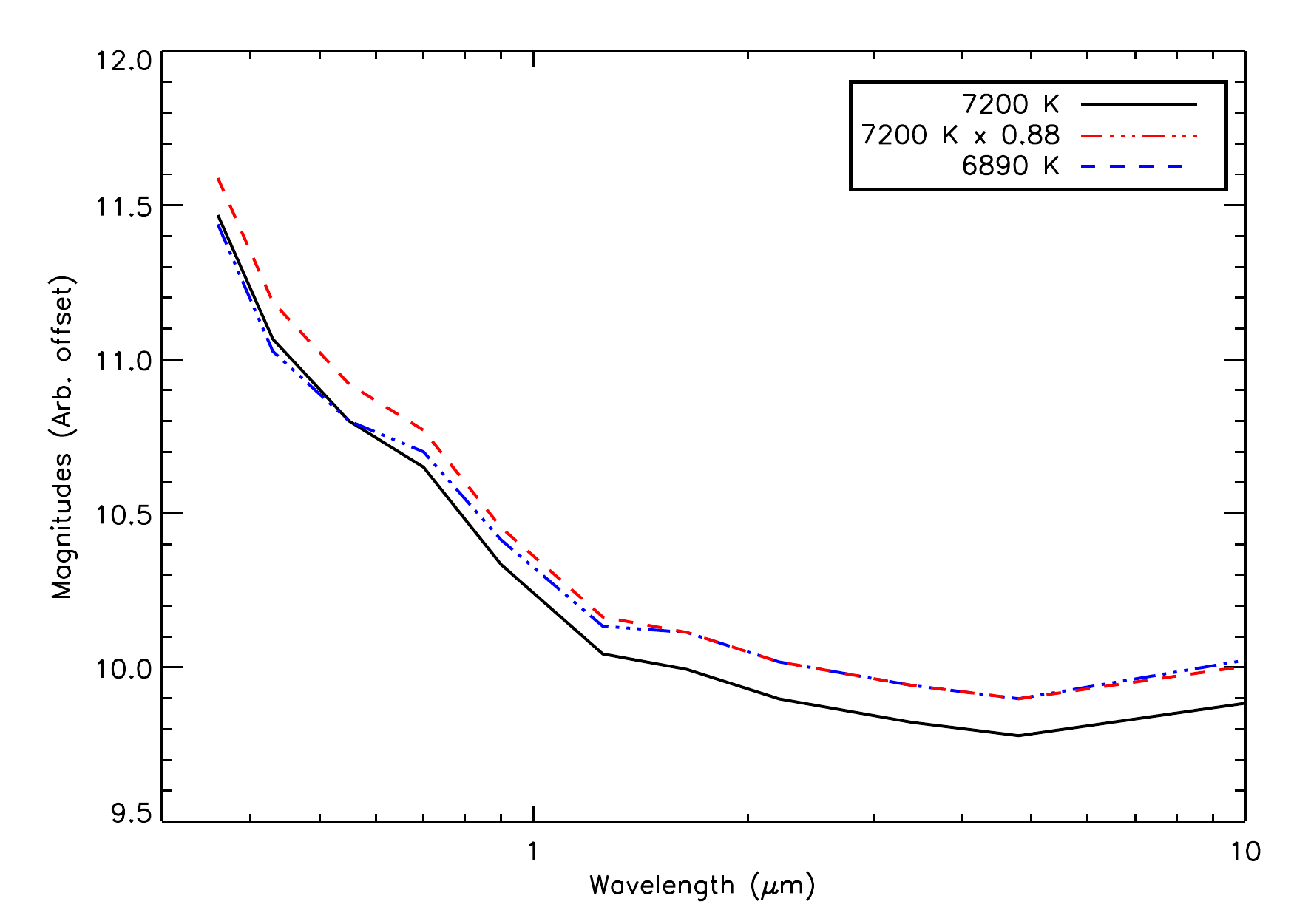}
\figcaption{\label{bbmodel} Three models using standard photometric colors of young stars (reddened to an A$_V$ of 1.8) from \citet{kenyon95}: 7200 K (black solid line); 6890 K (blue dot-dashed line); 7200 K decreased by 12\% (0.12 mag) (red dashed line).  The magnitude scale has an arbitrary offset.  Cooling from 7200 K to 6890 K is virtually indistinguishable from a 12\% decrease in flux at wavelengths $>$ 1 $\mu$m, but cooling would not result in a significant decrease in UBVR flux density.}
\end{figure*}

\begin{figure*}
\centering
\includegraphics[scale=0.6,trim={0 0 0 0},clip]{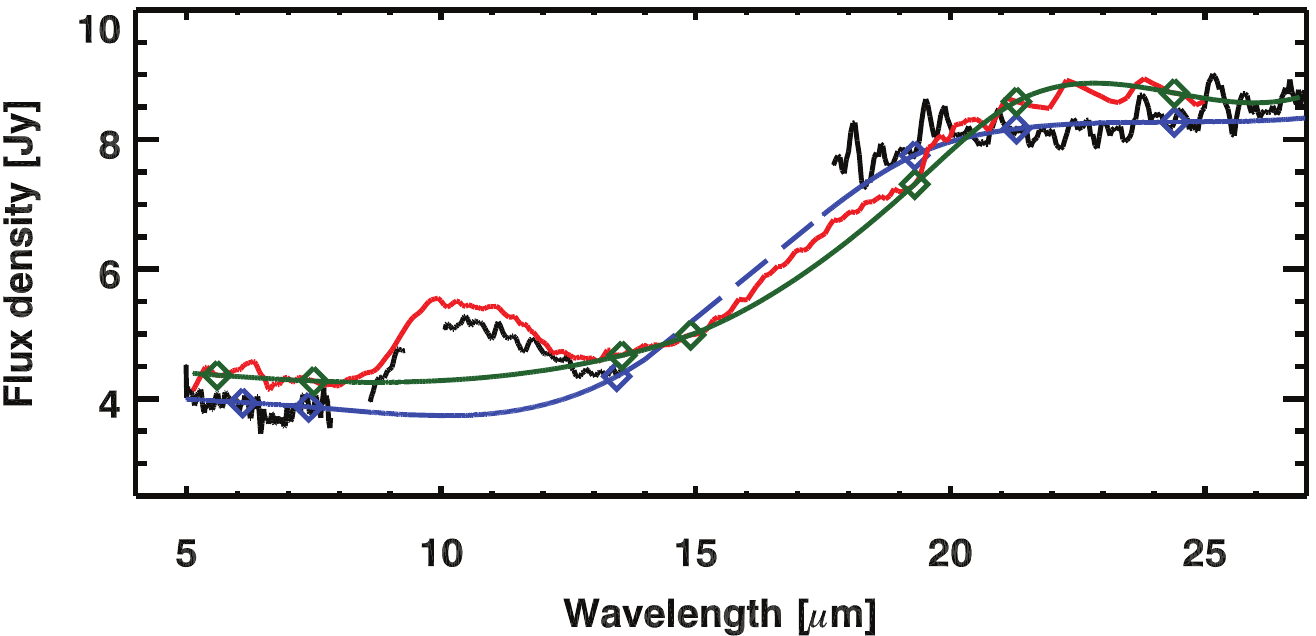}\includegraphics[scale=0.6,trim={0 0 0 0},clip]{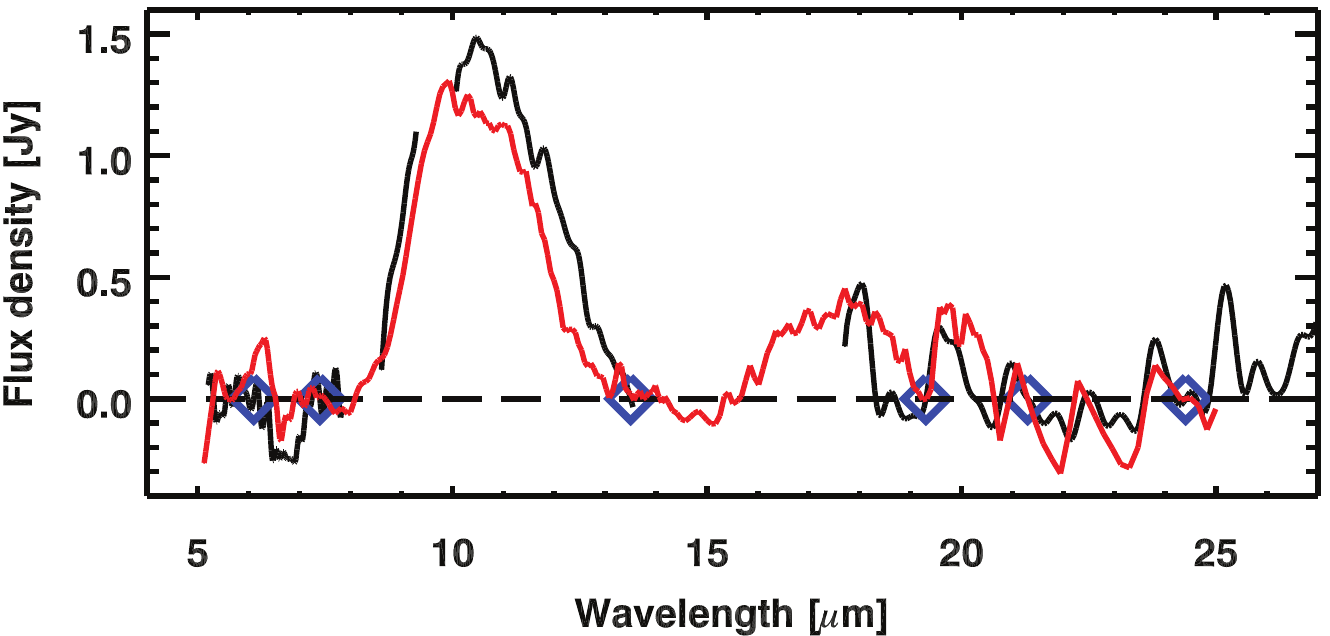}
\includegraphics[scale=0.6,trim={0 0 0 0},clip]{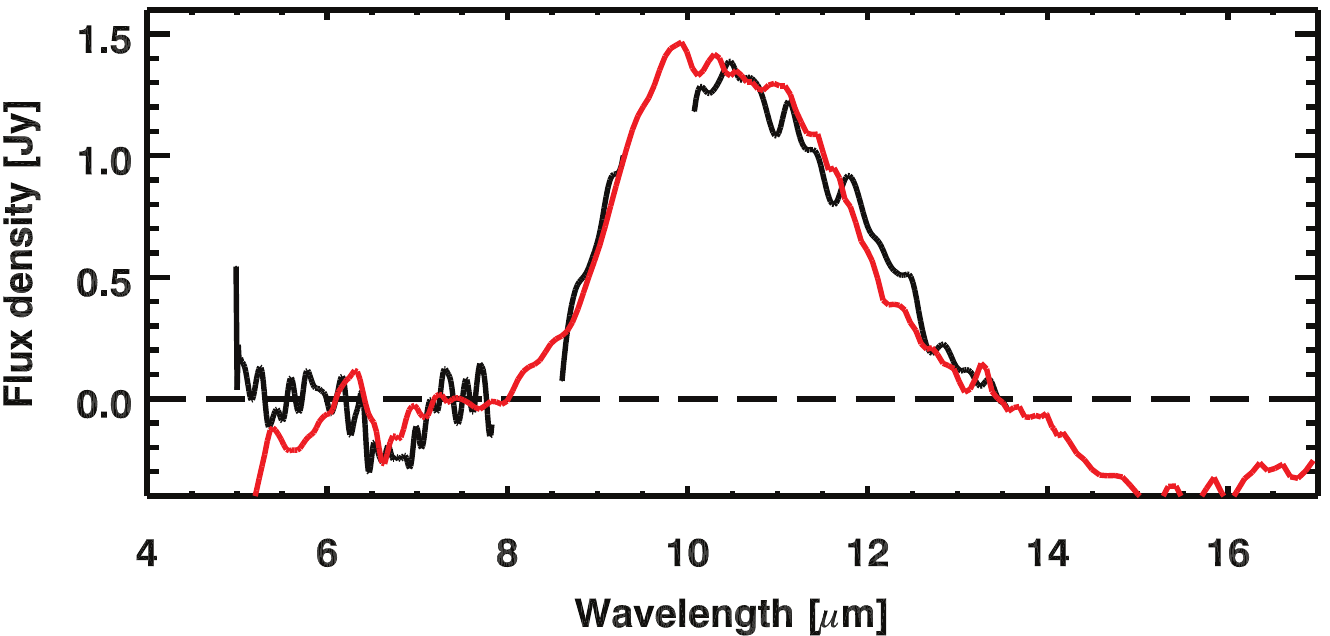}
\figcaption{\label{silicate} {\bf: Top Left:} Independent spline fits to the SOFIA and Spitzer spectra of FU Ori.  The IRS (red) spectrum is fit by the green solid spline, and the FORCAST (black) spectrum is fit by the blue spline, including an interpolation between 14 and 18 $\mu$m (blue dashed line). The spline anchor points are marked with blue and green diamonds.  The spline anchor points were determined independently for the two datasets (see text).   {\bf Top Right:} The resulting spline-fit subtracted IRS (red) and FORCAST (black) spectra.  {\bf Bottom:} Spline-fit subtracted IRS and FORCAST spectra with five {\it identical} spline anchor points, highlighting the consistency of the silicate feature (see text).}
\end{figure*}

\begin{figure*}
\centering
\includegraphics[scale=0.75,trim={0 0 0 0},clip]{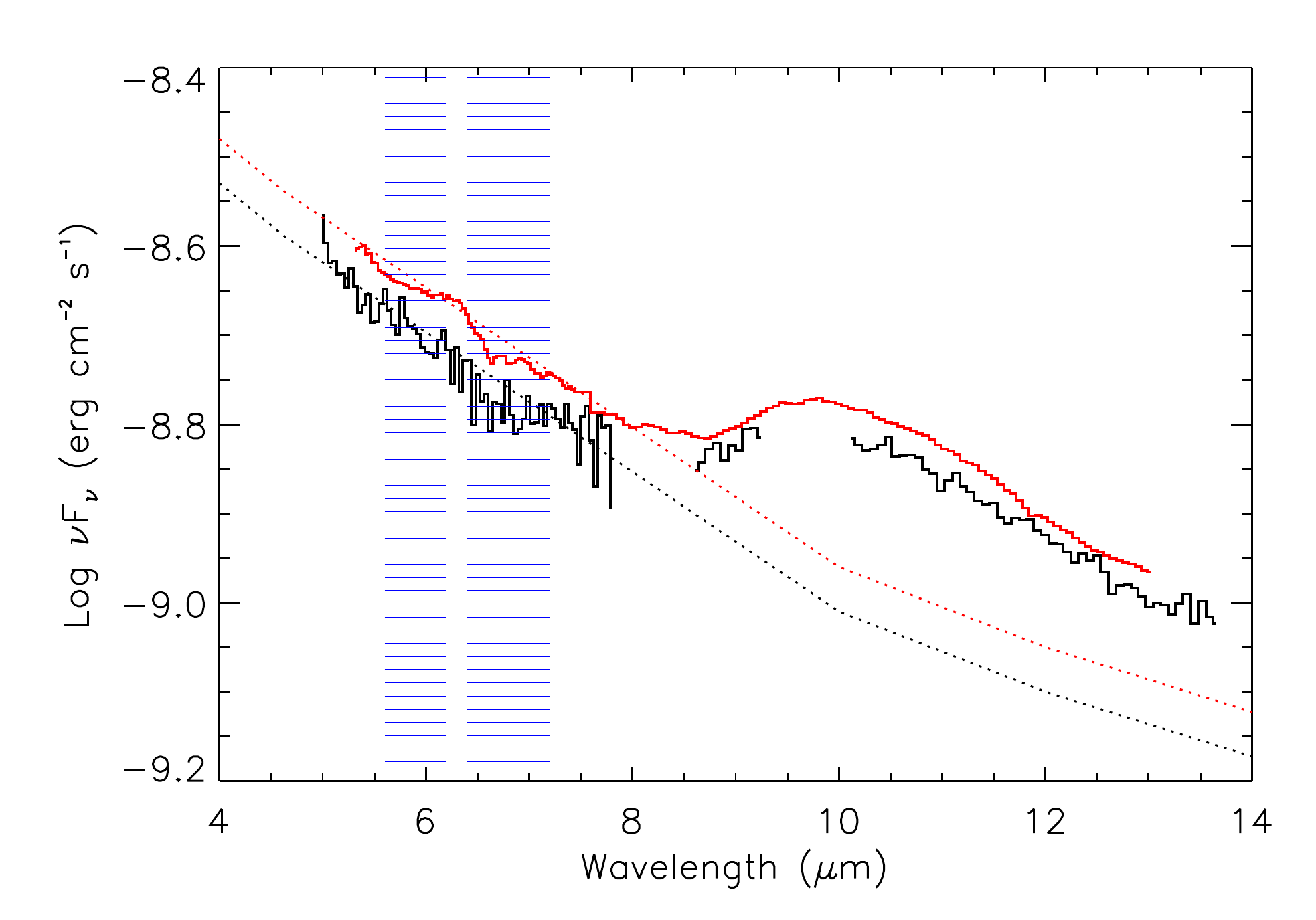}
\figcaption{\label{fig:model} Comparison of Spitzer-SL (red) and SOFIA G063 and G111 (black) spectra covering 5-14 $\mu$m, along with 7200 K modified blackbody models as in Figure \ref{fig:fit}.   The SOFIA spectrum has been smoothed to low resolution for easier comparison with the SL data.  The water absorption features can be clearly seen in both spectra, marked by the blue shaded regions \citep{gonzalez98}. }\end{figure*}

\begin{figure*}
\centering
\includegraphics[scale=0.75,trim={0 0 0 0},clip]{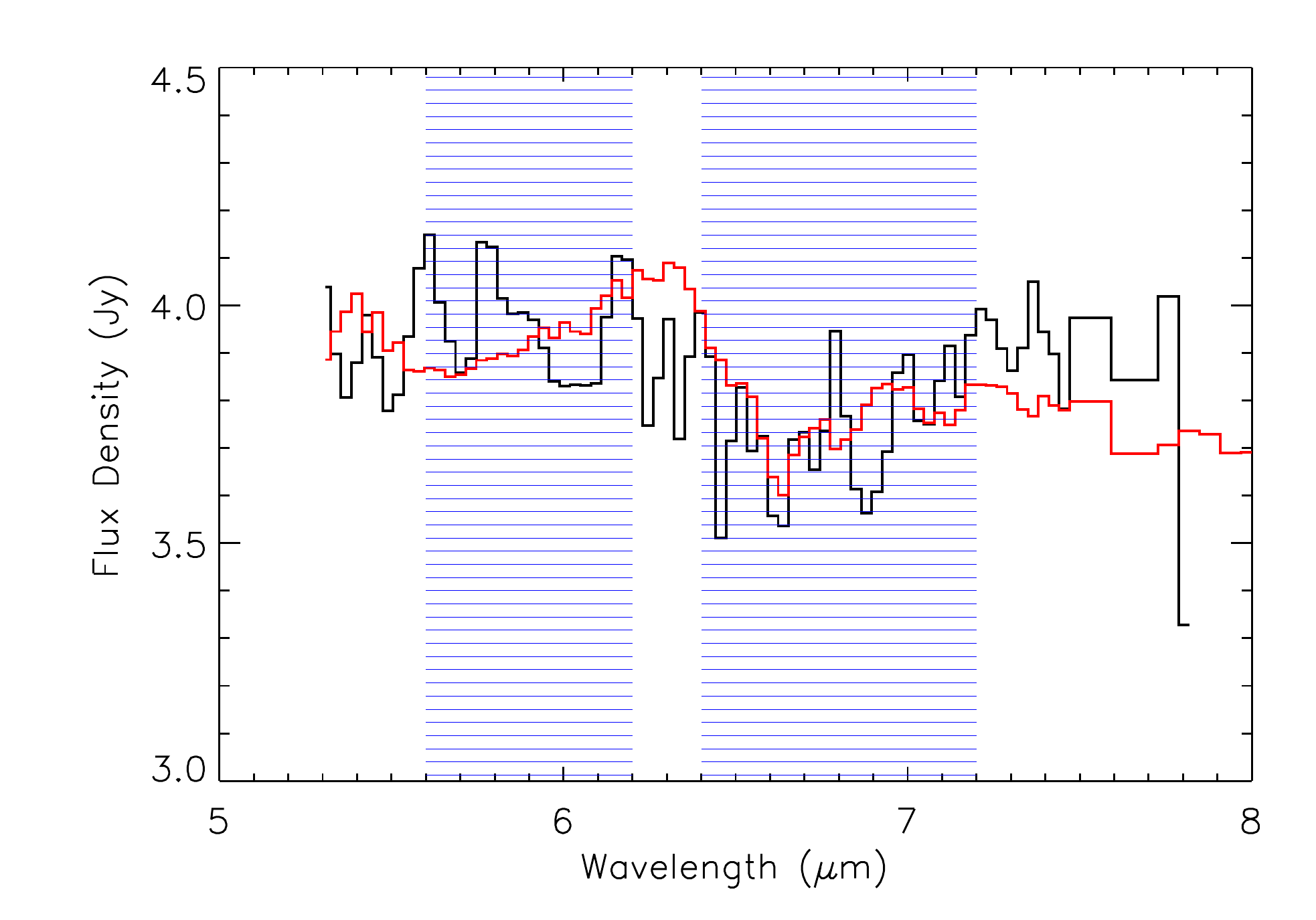}
\figcaption{\label{fig:model2} Zoom in of Spitzer-SL (red) and SOFIA G063 (black) spectra covering 5-8 $\mu$m.  The blue shaded regions indicate the wavelengths of the R- and P-branch H$_2$O lines \citep{gonzalez98}.    In this figure, the Spitzer spectrum has been offset -0.5 Jy to align with the SOFIA data.  The P-branch water absorption complex can be clearly seen in both spectra, but the R-branch lines appear weak in the SOFIA spectrum. }\end{figure*}

\section{Analysis}

\subsection{Continuum}

First we consider changes in the continuum flux density.    The (red) photometry presented in Figure \ref{fig:fit} comes from various times between 1998 and 2010, but we used the AAVSO UBVR and decline rate in B and V (0.015 mag yr$^{-1}$) measured for FU Ori in previous work \citep{kenyon00, green06} to confirm that FU Ori has continued to decline in B and V-band.  Specifically, we averaged the B and V magnitudes measured in 2016 from the AAVSO, 11.178 $\pm$ 0.078 and 9.844 $\pm$ 0.156, respectively. 
Photometry contemporaneous with the Spitzer spectroscopy \citep{green06}, taken in Oct 2004, indicated B and V magnitudes of 11.01 $\pm$ 0.02 and 9.66 $\pm$ 0.02, respectively.   This data is shown in Table \ref{photlog}. Extrapolating from Oct 2004 to Mar 2016 indicates decline rates of $\sim$ 0.015 and 0.016 mag/yr, respectively.   This translates to a decrease of 14-15\% in the optical bands over the 12 years between the Spitzer and SOFIA observations.

Next we consider the mid-IR bands by comparing the common portions of the 5-28 $\mu$m data from Spitzer and SOFIA directly; the ratio is shown in Figure \ref{ratio}.  We observe a $\sim$ 12\% decrease from 5-8$\mu$m, and a $\sim$ 7\% decrease from 10-28 $\mu$m; these ratios are shown in Figure \ref{ratio}.

Finally we consider the entire 0.1-100 $\mu$m spectrum.  We began by fitting the Spitzer 5-8 $\mu$m continuum, along with the UBVRIJHK, and WISE W1 band similar to \citet{green06}, using the modified blackbody models from \citet{kenyon95}.  We found that any of the 6590-7200 K models were equivalent fits to the observed 5-8 $\mu$m slope.  There are larger differences between the models near 1 $\mu$m, but lacking ca. 2016 data in the near-IR, the uncertainties are too great to distinguish between them.  For the purpose of this comparison, we adopt the 7200 K model as in \citet{green06}.  We then fit the rebinned SOFIA continuum at the same anchor wavelengths and found that the same 7200 K model, but with 12\% lower flux, matched the continuum to within uncertainties.

We found the slopes were not sufficiently distinguishable in either the optical/near-IR photometry or in the 5-8 $\mu$m spectroscopy. To get an idea of the uncertainties in the temperature, we used standard photometric colors from Table A.5 of \citet{kenyon95}.  To illustrate this, Figure \ref{bbmodel} shows three modified blackbody models reddened to an A$_V$ of 1.8 \citep{mathis90}, similar to FU Ori: 7200 K, 6890 K, and 7200 K but decreased by 12\%; the difference between cooling from 7200 to 6890 K, rather than decreasing the accretion rate by 12\%, is indistinguishable at wavelengths $>$ 1 $\mu$m. Given that the UBVR photometry we include shows a decrease comparable to the IR, it is most plausible to explain the continuum flux change by depletion of the innermost region of the disk (or equivalently decreased accretion rate); however, the mid-IR flux decrease can be explained by cooling (a decrease in the inner disk temperature). The cooling time in centrally irradiated T Tauri disks is very short, either a few years down to a few days depending upon the disk geometry \citep{dalessio99}.  Additionally,  for typical FUor system parameters, \citet{johnstone13} found a cooling time in FUor disks on the order of a few weeks (albeit long compared with the dust-photon coupling timescale).  Assuming the outer disk of FU Ori remains similar to a T Tauri disk, the cooling timescale is too rapid to explain the slow rate of decay.

Typically the draining of the gas within a disk would be noted in a decrease of the flux of the corotating molecular gas, as was seen in the post-outburst emission lines of EX Lup \citep{banzatti15}.  Unlike T Tauri disks \citep{salyk11}, the CO lines are seen in absorption in FUors, even at 4.6 $\mu$m \citep{smith09}; they attribute the absorption to a surrounding envelope of 0.1 M$_{\odot}$, which FU Ori does not possess \citep[e.g.][]{green13c}.

\subsection{Dust features}

The 10 $\mu$m silicate emission feature in the Spitzer-IRS spectrum showed no evidence of crystalline grains, but some evidence for grain growth during the disk lifetime \citep{quanz07a}.  In order to compare the SOFIA feature, we used a spline fit to anchor the spectrum at 8 $\mu$m, between the SiO and water vapor features, at 13 $\mu$m (the long wavelength edge of the G111 spectrum), and beyond 20 $\mu$m (centered within the G253 spectrum).  

Two different spline fits to both the Spitzer and SOFIA spectra are shown in Figure \ref{silicate}, and described below.  We first tried independent spline fitting on each spectrum: 5.6, 7.5, 13.6,14.9, 19.3, 21.3, and 24.4 $\mu$m anchor points for the Spitzer data; 6.1, 7.4, 13.5, 19.3, 21.3, and 24.4 $\mu$m for the SOFIA data.  We identified the anchor continuum points separately, and unlike the IRS spectrum, the FORCAST spectrum does not have continuous wavelength coverage from 5-38 $\mu$m, nor over the 9-13 $\mu$m complete silicate complex.  Thus the fits are slightly different, and the residuals appear to have shifted (Figure \ref{silicate}, top panel).  Specifically, the 10 $\mu$m silicate feature {\it appears} broader (to longer wavelengths) and taller.  To check whether this difference was due to the differences in spline fitting between the two spectra, we performed a second spline fitting and subtraction using exactly the same anchor points for both the SOFIA-FORCAST and the Spitzer-IRS spectrum.  Because we were limited to wavelength points common to both spectra, we used only six anchor points rather than the original seven in the IRS spectrum.  The results are shown in the bottom of Figure \ref{silicate}.  The features are virtually unchanged given the S/N of the SOFIA spectrum; we see no evidence of further dust processing or grain growth in the last 12 years, and attribute any apparent differences in the feature to incomplete spectral coverage in the FORCAST bands.  

Unfortunately, we lack sufficient spectral coverage to compare the 20 $\mu$m silicate features, but we find them consistent at wavelengths greater than 20 $\mu$m, and attribute any differences at shorter wavelengths to edge effects in the G253 spectrum.

\subsection{Gaseous absorption features: H$_2$O}

The blended rovibrational absorption lines of H$_2$O, originally noted in \citet{green06}, are still present in the SOFIA-FORCAST spectrum (Figure \ref{fig:model}).  The short wavelength shoulder marks the boundary between the blended R-branch lines ($\lambda$ $<$ 6.3 $\mu$m) and the blended P-branch lines ($\lambda$ $>$ 6.3 $\mu$m). The blended P-branch absorption lines appear unchanged over the last 12 years.  However, the relatively shallower blended R-branch absorption detected in the Spitzer-IRS spectrum is not clearly detected.  In order to compare the features directly, we apply a -0.5 Jy offset to the IRS spectrum, shown in Figure \ref{fig:model2}.  Differences between the R- and P-branch lines were noted, for example, in Orion-BN/KL \citep{gonzalez98}, who found the R-branch in absorption and P-branch in emission.  However, further observations with higher S/N are needed to confirm any changes.

Additional absorption attributed to SiO (8 $\mu$m) by \citet{green06} may be present in the FORCAST spectrum, but without full spectral coverage, or improved sensitivity, we cannot determine whether this feature has changed in detail.

\section{Discussion}

FUors are generally understood as large accretion bursts, although models differ on where the infalling material begins its descent. Binarity and outburst behavior has been examined in several studies \citep[e.g.][]{bonnell92,reipurth02,millan06,green16b}.  As noted earlier, FU Ori is itself a 0$\farcs$5 binary, and disk-binary interactions could promote an episodic or burst mode of accretion \citep{reipurth04}. Assuming the central star mass of FU Ori is $M \sim 0.3 M_{\odot}$ \citep{zhu09}, the integrated accretion luminosity (using model parameters from \citealt{green06}) corresponds to a total accreted mass of $M \sim 0.018 M_{\odot}$. Furthermore, FU Ori was the target of a recent ALMA study that separated the two components \citep{hales15} and demonstrated that the remnant envelope is tenuous at best, indicating that there is no external reservoir for resupplying the disk material in repeated bursts. They hypothesized that the more extinguished southern component could act as the mass reservoir for resupply.   Flattened envelope models can explain the long wavelength SED and provide long-term resupply potential, although whether sufficient mass exists for resupply is unclear, judging by the far-IR excess reported by \citet{green13c}, who estimated a dust mass of 0.05  M$_{\odot}$ from the Herschel-SPIRE $>$ 350 $\mu$m excess.  Higher resolution observations from the Submillimeter Array further contained the size and mass of such an envelope plus disk observations to 0.05 M$_{\odot}$ within 450 AU or less (M. Dunham, 2016, priv. comm., \citealt{green16b}).

Regardless of the origin of the infalling material, our study suggests that at the mature stages of outburst, the infalling material accretes from the inside-out.  The decrease in only the short wavelength continuum, without change to spectral shape, indicates a decrease in density of the hot material, rather than a decrease in temperature.  Combined with the lack of change in the silicate dust features after removing continuum effects, we suggest that the luminosity decrease in FU Ori is due primarily to this draining effect, leading to a decrease in the accretion rate.  Because the peak temperature in the disk still appears to be about 7200 K, the infall radius (centrifugal radius) is quite close to the central star.  \citet{zhu07} modeled FU Ori as a superheated disk with a maximum effective temperature of 6420 K at $r_{in}$ $=$ 5 R$_{\odot}$, decreasing to 4660 K at 3 $r_{in}$, eventually to 977 K at 30 $r_{in}$ (0.7 AU).  They added a second component from a more typical accretion disk $>$ 1 AU.  

Supposing over the past 12 years FU Ori accreted only material at a temperature $>$ 977 K, the accreted material would all be within $\sim$ 0.7 AU, where the free fall time is much shorter than 12 years.  It may be refilled from surrounding regions but the maximum refill rate is capped by the upper limit to the change in IR excess beyond 20 $\mu$m, which arises from annuli beyond the boundaries of the superheated disk at $\sim$ 1 AU \citep{zhu09}.   It appears that the outer disk does not react as quickly to this decreased luminosity on a 12 year timescale; within the uncertainties in scatter, the continuum level $<$ 20 $\mu$m has changed by $\sim$ 7\%.  

However, we note that we cannot rule out cooling in the inner disk from this change in the SED.  If this is due to cooling of the inner disk rather than a decrease in the accretion rate, and the 5-8 $\mu$m is dominated by the hot disk as models suggest, then the temperature decrease from a 12\% decrease in luminosity would be only 4\% (ie. from 7200 K to 6900 K).  This is a small effect on the slope and is consistent with our data in the mid-IR (Figure \ref{bbmodel}).  Potentially, a cooling pattern and decreased central accretion could be distinguished with high fidelity IR continuum measurements, increased coverage in the optical-near IR bands, or by changes in the width or shape of CO lines from differing disk annuli. \citet{bell95} noted that the decrease in the SED of FUor V1057 Cyg was least significant between 1 and 10 $\mu$m; between 2004 and 2016, at least, FU Ori does not seem to follow this pattern, maintaining a decay rate similar to the optical bands out to at least 8 $\mu$m.  This difference may reflect the relatively small amount of material surrounding FU Ori, or the overall slower decay rate compared with that of V1057 Cyg.  An alternate explanation is that the FU Ori disk is geometrically thicker due to the extreme nature and duration of the accretion event (e.g., \citealt{hartmann11} and references therein), and is beginning to settle as the midplane begins to cool.

\section{Conclusions}

We have performed the first comparison of multi-epoch mid-IR spectra of an FUor, to examine the changes occurring in the system during outburst.  We compared the ca. 2004 SED and IR spectrum of FU Orionis with newly acquired (2016) SOFIA-FORCAST spectrum and photometry to identify changes over the intervening 12 year period.  After establishing cross-calibration between the various measurements, we find a 12\% ($\sim$ 3$\sigma$) decrease in the 5-14 $\mu$m  continuum without significant change in shape.  We find lesser changes ($\lesssim$ 7\%) to the SED beyond 10 $\mu$m, and no significant change to the silicate dust or the rovibrational water absorption from the disk photosphere.  From this, we posit  a decrease in gas at the highest temperatures, and therefore the innermost regions of the disk, either due to cooling or depletion via accretion processes.

\acknowledgements

We thank L. Andrew Helton, Sachin Shenoy, James DeBuizer, William Vacca, Joseph Adams, Paule Sonnentrucker, Andrea Banzatti, Miguel Requena-Torres, Klaus Pontoppidan, and Alexander Hubbard for helpful discussions of our results.  We thank the SOFIA Science Center (USRA) for data reduction and analysis support.  This work is based in part on observations made with the NASA/DLR Stratospheric Observatory for Infrared Astronomy (SOFIA). SOFIA is jointly operated by the Universities Space Research Association, Inc. (USRA), under NASA contract NAS2-97001, and the Deutsches SOFIA Institut (DSI) under DLR contract 50 OK 0901 to the University of Stuttgart. Partial support for this work was provided by NASA through award \#NAS2-97001, SOF 04-0146 issued by USRA.  We acknowledge the variable star observations from the AAVSO International Database contributed by observers worldwide and used in this work, and in particular observer ``DUBF'' Franky Dubois. The work of WJF was supported by an appointment to the NASA Postdoctoral Program at Goddard Space Flight Center, administered by Universities Space Research Association under contract with NASA. OCJ acknowledges NASA grant, NNX14AN06G.  This research has made use of NASA's Astrophysics Data System Bibliographic Services. This research has made use of the VizieR catalogue access tool, CDS, Strasbourg, France.

\bibliographystyle{aasjournal}
\bibliography{dissbib}
\include{dissbib}

\end{document}